\documentstyle[twoside,psfig]{article}
\catcode`\@=11
\long\def\@makefntext#1{
\protect\noindent \hbox to 3.2pt {\hskip-.9pt  
$^{{\eightrm\@thefnmark}}$\hfil}#1\hfill}		

\def\@makefnmark{\hbox to 0pt{$^{\@thefnmark}$\hss}}	
	
\def\ps@myheadings{\let\@mkboth\@gobbletwo
\def\@oddhead{\hbox{}
\rightmark\hfil\eightrm\thepage}   
\def\@oddfoot{}\def\@evenhead{\eightrm\thepage\hfil
\leftmark\hbox{}}\def\@evenfoot{}
\def\sectionmark##1{}\def\subsectionmark##1{}}

\oddsidemargin=\evensidemargin
\newcounter{sectionc}\newcounter{subsectionc}\newcounter{subsubsectionc}
\renewcommand{\section}[1] {\vspace{12pt}\addtocounter{sectionc}{1} 
\setcounter{subsectionc}{0}\setcounter{subsubsectionc}{0}\noindent 
	{\tenbf\thesectionc. #1}\par\vspace{5pt}}
\renewcommand{\subsection}[1] {\vspace{12pt}\addtocounter{subsectionc}{1} 
	\setcounter{subsubsectionc}{0}\noindent 
	{\bf\thesectionc.\thesubsectionc. {\kern1pt \bfit #1}}\par\vspace{5pt}}
\renewcommand{\subsubsection}[1] {\vspace{12pt}\addtocounter{subsubsectionc}{1}
	\noindent{\tenrm\thesectionc.\thesubsectionc.\thesubsubsectionc.
	{\kern1pt \tenit #1}}\par\vspace{5pt}}
\newcommand{\nonumsection}[1] {\vspace{12pt}\noindent{\tenbf #1}
	\par\vspace{5pt}}

\newcounter{appendixc}
\newcounter{subappendixc}[appendixc]
\newcounter{subsubappendixc}[subappendixc]
\renewcommand{\thesubappendixc}{\Alph{appendixc}.\arabic{subappendixc}}
\renewcommand{\thesubsubappendixc}
	{\Alph{appendixc}.\arabic{subappendixc}.\arabic{subsubappendixc}}

\renewcommand{\appendix}[1] {\vspace{12pt}
        \refstepcounter{appendixc}
        \setcounter{figure}{0}
        \setcounter{table}{0}
        \setcounter{lemma}{0}
        \setcounter{theorem}{0}
        \setcounter{corollary}{0}
        \setcounter{definition}{0}
        \setcounter{equation}{0}
        \renewcommand{\thefigure}{\Alph{appendixc}.\arabic{figure}}
        \renewcommand{\thetable}{\Alph{appendixc}.\arabic{table}}
        \renewcommand{\theappendixc}{\Alph{appendixc}}
        \renewcommand{\thelemma}{\Alph{appendixc}.\arabic{lemma}}
        \renewcommand{\thetheorem}{\Alph{appendixc}.\arabic{theorem}}
        \renewcommand{\thedefinition}{\Alph{appendixc}.\arabic{definition}}
        \renewcommand{\thecorollary}{\Alph{appendixc}.\arabic{corollary}}
        \renewcommand{\theequation}{\Alph{appendixc}.\arabic{equation}}
        \noindent{\tenbf Appendix \theappendixc #1}\par\vspace{5pt}}
\newcommand{\subappendix}[1] {\vspace{12pt}
        \refstepcounter{subappendixc}
        \noindent{\bf Appendix \thesubappendixc. {\kern1pt \bfit #1}}
	\par\vspace{5pt}}
\newcommand{\subsubappendix}[1] {\vspace{12pt}
        \refstepcounter{subsubappendixc}
        \noindent{\rm Appendix \thesubsubappendixc. {\kern1pt \tenit #1}}
	\par\vspace{5pt}}

\newcommand{\textlineskip}{\baselineskip=13pt}
\newcommand{\smalllineskip}{\baselineskip=10pt}

\def\eightcirc{
\begin{picture}(0,0)
\put(4.4,1.8){\circle{6.5}}
\end{picture}}
\def\eightcopyright{\eightcirc\kern2.7pt\hbox{\eightrm c}} 

\newcommand{\copyrightheading}[1]
	{\vspace*{-2.5cm}\smalllineskip{\flushleft
	{\footnotesize Modern Physics Letters A, #1}\\
	{\footnotesize $\eightcopyright$\, World Scientific Publishing
	 Company}\\
	 }}

\def\abstracts#1#2#3{{
	\centering{\begin{minipage}{4.5in}\footnotesize\baselineskip=10pt
	\parindent=0pt #1\par 
	\parindent=15pt #2\par
	\parindent=15pt #3
	\end{minipage}}\par}} 

\def\keywords#1{{
	\centering{\begin{minipage}{4.5in}\footnotesize\baselineskip=10pt
	{\footnotesize\it Keywords}\/: #1
	 \end{minipage}}\par}}

\renewenvironment{thebibliography}[1]
	{\frenchspacing
	 \ninerm\baselineskip=11pt
	 \begin{list}{\arabic{enumi}.}
        {\usecounter{enumi}\setlength{\parsep}{0pt}     
	 \setlength{\leftmargin 12.7pt}{\rightmargin 0pt} 
         \setlength{\itemsep}{0pt} \settowidth
	{\labelwidth}{#1.}\sloppy}}{\end{list}}

\newcounter{itemlistc}
\newcounter{romanlistc}
\newcounter{alphlistc}
\newcounter{arabiclistc}

\newcommand{\fcaption}[1]{
        \refstepcounter{figure}
        \setbox\@tempboxa = \hbox{\footnotesize Fig.~\thefigure. #1}
        \ifdim \wd\@tempboxa > 5in
           {\begin{center}
        \parbox{5in}{\footnotesize\smalllineskip Fig.~\thefigure. #1}
            \end{center}}
        \else
             {\begin{center}
             {\footnotesize Fig.~\thefigure. #1}
              \end{center}}
        \fi}

\newcommand{\tcaption}[1]{
        \refstepcounter{table}
        \setbox\@tempboxa = \hbox{\footnotesize Table~\thetable. #1}
        \ifdim \wd\@tempboxa > 5in
           {\begin{center}
        \parbox{5in}{\footnotesize\smalllineskip Table~\thetable. #1}
            \end{center}}
        \else
             {\begin{center}
             {\footnotesize Table~\thetable. #1}
              \end{center}}
        \fi}

\def\@citex[#1]#2{\if@filesw\immediate\write\@auxout
	{\string\citation{#2}}\fi
\def\@citea{}\@cite{\@for\@citeb:=#2\do
	{\@citea\def\@citea{,}\@ifundefined
	{b@\@citeb}{{\bf ?}\@warning
	{Citation `\@citeb' on page \thepage \space undefined}}
	{\csname b@\@citeb\endcsname}}}{#1}}

\newif\if@cghi
\def\cite{\@cghitrue\@ifnextchar [{\@tempswatrue
	\@citex}{\@tempswafalse\@citex[]}}
\def\citelow{\@cghifalse\@ifnextchar [{\@tempswatrue
	\@citex}{\@tempswafalse\@citex[]}}
\def\@cite#1#2{{$\null^{#1}$\if@tempswa\typeout
	{IJCGA warning: optional citation argument 
	ignored: `#2'} \fi}}

\def\pmb#1{\setbox0=\hbox{#1}
	\kern-.025em\copy0\kern-\wd0
	\kern.05em\copy0\kern-\wd0
	\kern-.025em\raise.0433em\box0}

\def\fnt#1#2{\footnotetext{\kern-.3em
	{$^{\mbox{\scriptsize #1}}$}{#2}}}

\def\fpage#1{\begingroup
\voffset=.3in
\thispagestyle{empty}\begin{table}[b]\centerline{\footnotesize #1}
	\end{table}\endgroup}

\def\runninghead#1#2{\pagestyle{myheadings}
\markboth{{\protect\footnotesize\it{\quad #1}}\hfill}
{\hfill{\protect\footnotesize\it{#2\quad}}}}
\font\tenrm=cmr10
\font\tenit=cmti10 
\font\tenbf=cmbx10
\font\bfit=cmbxti10 at 10pt
\font\ninerm=cmr9

\font\eightrm=cmr8

\def\qed{\hbox{${\vcenter{\vbox{			
   \hrule height 0.4pt\hbox{\vrule width 0.4pt height 6pt
   \kern5pt\vrule width 0.4pt}\hrule height 0.4pt}}}$}}

\def\nc{\newcommand}
\nc{\bA}{\mbox{\boldmath $A$\unboldmath}}
\nc{\bn}{\mbox{\boldmath $n$\unboldmath}}
\nc{\bl}{\mbox{\boldmath $l$\unboldmath}}
\nc{\bm}{\mbox{\boldmath $m$\unboldmath}}

\nc{\cD}{\cal D}
\nc{\cL}{\cal L}
\nc{\cLd}{{\cal L}^{\dagger}}
\nc{\tkr}{2\kappa (r-r_H)}
\nc{\sta}{\sin\theta}
\nc{\cta}{\cos\theta}
\nc{\sda}{\sin^2\theta}
\nc{\cda}{\cos^2\theta}
\nc{\coa}{\cot\theta}
\nc{\sqd}{\sqrt{2}}

\def\p{\partial}
\nc{\pr}{\frac{\p}{\p r}}
\nc{\pv}{\frac{\p}{\p v}} 
\nc{\pta}{\frac{\p}{\p \theta}}
\nc{\pvi}{\frac{\p}{\p \varphi}}
\nc{\pdr}{\frac{\p^2}{\p r^2}}
\nc{\pdta}{\frac{\p^2}{\p \theta^2}}
\nc{\pdvi}{\frac{\p^2}{\p \varphi^2}}
\nc{\pdvr}{\frac{\p^2}{\p v \p r}}

\nc{\spr}{\frac{\p}{\p r_*}}
\nc{\spv}{\frac{\p}{\p v_*}}
\nc{\spdr}{\frac{\p^2}{\p r_*^2}}
\nc{\spdvr}{\frac{\p^2}{\p r_* \p v_*}}

\begin{document}
\setlength{\textheight}{7.7truein}  

\runninghead{Non-existence of New Quantum Ergosphere
Effect of a Vaidya-type Black Hole}{Wu and Cai}

\normalsize\textlineskip
\thispagestyle{empty}
\setcounter{page}{1}
\copyrightheading{Vol. , No.  (2001)  -- }
\vspace*{0.88truein}
\fpage{1}

\centerline{\bf Non-existence of New Quantum Ergosphere}
\baselineskip=13pt
\centerline{\bf Effect of a Vaidya-type Black Hole}
\vspace*{0.37truein}

\centerline{\footnotesize S. Q. Wu}
\baselineskip=12pt
\centerline{\footnotesize\it Institute of Particle Physics, 
Hua-Zhong Normal University}
\baselineskip=10pt
\centerline{\footnotesize\it Wuhan 430079, P.R. China}
\baselineskip=10pt
\centerline{\footnotesize\it E-mail: sqwu@iopp.ccnu.edu.cn}
\baselineskip=14pt

\centerline{\footnotesize X. Cai}
\baselineskip=12pt
\centerline{\footnotesize\it Institute of Particle Physics, 
Hua-Zhong Normal University}
\baselineskip=10pt
\centerline{\footnotesize\it Wuhan 430079, P.R. China}
\baselineskip=10pt
\centerline{\footnotesize\it E-mail: xcai@ccnu.edu.cn}
\vspace*{0.225truein}

\vspace*{0.21truein}
\abstracts{{\it Abstract}:
Hawking evaporation of Dirac particles and scalar fields 
in a Vaidya-type black hole is investigated by the method 
of generalized tortoise coordinate transformation. It is 
shown that Hawking radiation of Dirac particles does not 
exist for $P_1, Q_2$ components but for $P_2, Q_1$ components 
in any  Vaidya-type black holes. Both the location and the 
temperature of the event horizon change with time. The thermal 
radiation spectrum of Dirac particles is the same as that of 
Klein-Gordon particles. We demonstrates that there is no new 
quantum ergosphere effect in the thermal radiation of Dirac 
particles in any spherically symmetry black holes.}{}

\vspace*{10pt}

\vspace*{1pt}
\textlineskip
\section{Introduction}
\vspace*{-0.5pt}

Hawking's investigation of quantum effects\cite{Hawk} interpreted
as the emission of a thermal spectrum of particles by a black hole
event horizon sets a remarkable landmark in black hole physics.
During the last decade, the Hawking radiation of Dirac particles 
had been extensively investigated in some spherically symmetric 
and non-static black holes.\cite{Zhetc} However, most of these 
studies concentrated on the spin state $p=1/2$ of the four-component 
Dirac spinors. Recently, the Hawking radiation of Dirac particles 
of spin state $p = -1/2$ attracted a little more attention.\cite{LZ,LLM} 
In several papers, Li et al.\cite{LZ,LLM} declared that they had 
found a kind of new quantum thermal effect for the Vaidya-Bonner-de 
Sitter black hole. Basing upon the generalized Teukolsky-type master 
equation\cite{Teuk} for fields of spin ($s = 0, 1/2, 1$ and $2$ 
for the scalar, Dirac, electromagnetic and gravitational field, 
respectively) in a Vaidya-type space-time, they showed that 
the massive Dirac field of spin state $p = 1/2$ differs greatly 
from that of spin state $p = -1/2$ in radiative mechanism, and 
suggested that it originate from the variance of Dirac vacuum 
near the event horizon in the non-static space-times caused by 
spin state. Further, they conjectured that this effect originates 
from the quantum ergosphere,\cite{York} that is, the quantum 
ergosphere can influence the radiative mechanism of a black hole. 
As far as their master equations\cite{LZ,LLM} are concerned, 
their argument, in fact, sounds only for the massive spin-$1/2$ 
particles because only the mass of Klein-Gordon particle and that 
of Dirac particle are nonzero. An exotic feature of this effect is 
its obvious dependence on the mass $\mu_0$ of spin-$1/2$ particles.

In this letter, we study the Hawking effect of Dirac and Klein-Gordon 
particles in a Vaidya-type black hole by means of the generalized 
tortoise transformation (GTCT) method. We consider simultaneously the 
limiting forms of the first-order and second-order forms of 
Dirac equation near the event horizon because the Dirac spinors 
should satisfy both of them. The event horizon equation, the 
Hawking temperature and the thermal radiation spectrum of 
electrons are in accord with others. We prove rigorously 
that the Hawking radiation takes place only for $P_2, Q_1$ 
but not for $P_1, Q_2$ components of Dirac spinors. The origin 
of this asymmetry of the Hawking radiation of different spinorial 
components probably stem from the asymmetry of space-time in 
the advanced Eddington-Finkelstein coordinate system. Besides,
we point out that there could not have been any new quantum 
ergosphere effect in the Hawking radiation of Dirac particles 
in any spherically symmetric black hole whether it is static 
or non-static. This conclusion is contrary completely to that 
of Li's\cite{LZ,LLM} who argued that the radiative mechanism of 
massive spin fields depends on the spin state. 

The paper is outlined as follows: In section 2, the explicit form 
of Dirac equation in the Vaidya-type black hole is presented in 
spinorial formalism, the event horizon equation is derived in 
Sec. 3. Then we obtain the Hawking temperature and the thermal 
radiation spectrum in Sec. 4 and 5, respectively. The Hawking
evaporation of Klein-Gordon particles is re-examined in Sec. 6. 
Finally we give some discussions.

\section{Dirac equation}

The metric of a Vaidya-type black hole with the 
cosmological constant $\Lambda$ is given in the 
advanced Eddington-Finkelstein coordinate system by 
\begin{equation}
ds^2 = 2dv(G dv -dr) -r^2(d\theta^2 
+\sin^2\theta d\varphi^2) \, , 
\end{equation}
where $2G = 1 -\frac{2M(v)}{r} -\frac{\Lambda}{3}r^4$, 
in which the mass $M$ of the hole is a function 
of the advanced time $v$.

We establish such a complex null-tetrad 
$\{\bl, \bn, \bm, \overline{\bm}\}$  
that satisfies the orthogonal conditions 
$\bl \cdot \bn = -\bm \cdot \overline{\bm} = 1$. 
Thus the covariant one-forms can be written as
\begin{eqnarray}
&&\bl = dv \, ,  ~~~~~~~~~~~~
\bm = \frac{-r}{\sqd}\left(d\theta 
+i\sta d\varphi\right) \, , \nonumber\\
&&\bn = G dv -dr \, , ~~ 
\overline{\bm} = \frac{-r}{\sqd}\left(d\theta 
-i\sta d\varphi\right) \, . 
\end{eqnarray}
and their corresponding directional derivatives are
\begin{eqnarray}
&&D = -\pr \, ,  ~~~~~~~~~~
\delta = \frac{1}{\sqd r}\left(\pta 
+\frac{i}{\sta}\pvi\right) \, , \nonumber \\ 
&&\Delta = \pv +G\pr \, , ~~
\overline{\delta} = \frac{1}{\sqd r}
\left(\pta -\frac{i}{\sta}\pvi\right) \, .  
\end{eqnarray}
It is not difficult to compute the non-vanishing 
Newman-Penrose complex spin coefficients\cite{NP} 
in the above null-tetrad as follows 
\begin{equation}
\mu = \frac{G}{r} \, , 
~~\gamma = -\frac{G_{,r}}{2} = -\frac{dG}{2dr} \, , 
~~\beta = -\alpha = \frac{\coa}{2\sqd r}  \, .  
\end{equation}

Inserting for the needed Newman-Penrose spin coefficients 
into the spinor form of the four coupled Chandrasekhar-Dirac 
equations\cite{CD} describing the dynamic behavior of 
spin-$1/2$ particles,
\begin{eqnarray}
&&(D +\epsilon -\rho)F_1 +(\overline{\delta} +\tilde{\pi} 
-\alpha)F_2 =\frac{i\mu_0}{\sqd}G_1 \, , \nonumber \\
&&(\Delta +\mu -\gamma)F_2 +(\delta +\beta -\tau)F_1 
=\frac{i\mu_0}{\sqd}G_2 \, , \nonumber\\
&&(D +\epsilon^* -\rho^*)G_2 -(\delta +\tilde{\pi}^*-\alpha^*)G_1 
=\frac{i\mu_0}{\sqd}F_2 \, ,  \nonumber\\
&&(\Delta +\mu^* -\gamma^*)G_1 -(\overline{\delta}
+\beta^* -\tau^*)G_2 =\frac{i\mu_0}{\sqd}F_1 \, ,
\end{eqnarray}
where $\mu_0$ is the mass of Dirac particles, one arrives at 
\begin{eqnarray}
-\left(\pr + \frac{1}{r}\right)F_1 +\frac{1}{\sqd r} {\cL}_{1/2} F_2 
= \frac{i\mu_0}{\sqd} G_1 \, ,  \nonumber&&\\
\frac{1}{2r^2} {\cD} F_2  +\frac{1}{\sqd r} {\cLd}_{1/2} F_1 
= \frac{i\mu_0}{\sqd} G_2 \, ,  \nonumber&&\\
-\left(\pr + \frac{1}{r}\right)G_2 -\frac{1}{\sqd r} {\cLd}_{1/2} G_1 
= \frac{i\mu_0}{\sqd} F_2 \, ,  \nonumber&&\\
\frac{1}{2r^2} {\cD} G_1 -\frac{1}{\sqd r} {\cL}_{1/2} G_2 
= \frac{i\mu_0}{\sqd} F_1 \, , \label{DCP}&&
\end{eqnarray}
in which operators
\begin{eqnarray*}
&&{\cD} = 2r^2\left(\pv +G\pr \right) +(r^2G)_{,r} \, ,\\
&&{\cL}_{1/2} = \pta +\frac{1}{2}\coa -\frac{i}{\sta}\pvi  \, , \\
&&{\cLd}_{1/2} = \pta +\frac{1}{2}\coa +\frac{i}{\sta}\pvi  \, .
\end{eqnarray*}
have been defined.

Further substitutions
$P_1 = \sqd r F_1, P_2 = F_2, 
Q_1 = G_1, Q_2 = \sqd r Q_2$ 
yields
\begin{eqnarray}
&&-\pr P_1 +{\cL}_{1/2} P_2 = i\mu_0 r Q_1 \, , ~~
{\cD} P_2 +{\cLd}_{1/2} P_1 = i\mu_0 r Q_2 \, ,  \nonumber\\
&&-\pr Q_2 -{\cLd}_{1/2} Q_1 = i\mu_0 r P_2 \, , ~~
{\cD} Q_1 -{\cL}_{1/2} Q_2 = i\mu_0 r P_1 \, . 
\label{reDP}
\end{eqnarray}
An apparent fact is that the Chandrasekhar-Dirac equation 
(\ref{reDP}) could be satisfied by identifying $Q_1$, $Q_2$ 
with $P_2^*$, $-P_1^*$, respectively. So one may deal with 
a pair of components $P_1$, $P_2$ only. 

\section{Event Horizon}

Now separating variables to Eq. (\ref{reDP}) as
$$P_1 = R_1(v,r)S_1(\theta,\varphi) \, , 
~~P_2 = R_2(v,r)S_2(\theta,\varphi) \, , $$
$$Q_1 = R_2(v,r)S_1(\theta,\varphi) \, , 
~~Q_2 = R_1(v,r)S_2(\theta,\varphi) \, , $$
then we can decouple it to the radial part
\begin{equation}
\pr R_1 = (\lambda - i\mu_0r) R_2 \, , ~~
{\cD} R_2 = (\lambda + i\mu_0r) R_1 \, , \label{sepa}
\end{equation}
and the angular part
\begin{equation}
{\cLd}_{1/2} S_1 = -\lambda S_2 \, , 
~~{\cL}_{1/2} S_2 = \lambda S_1 \, ,
\end{equation}
where $\lambda = \ell +1/2$ is a separation constant. 
Both functions $S_1(\theta,\varphi)$ and $S_2(\theta,\varphi)$ 
are, respectively, spinorial spherical harmonics 
$_sY_{\ell m}(\theta,\varphi)$ with spin-weight 
$s = \pm 1/2$ satisfying\cite{GMNRS}
\begin{eqnarray}
&&\Big[\pdta +\coa \pta +\frac{1}{\sda}\pdvi 
+\frac{2is\cta}{\sda}\pvi\nonumber \\ 
&&~~~~~~ -s^2\cot^2\theta +s +(\ell -s)(\ell +s +1)\Big] 
{_sY}_{\ell m}(\theta,\varphi) = 0 \, .   
\end{eqnarray}

As to the thermal radiation, one should be concerned 
about the behavior of the radial part of Eq. (\ref{sepa}) 
near the horizon only. Because the Vaidya-type black hole 
is spherically symmetric, one can introduce as a working 
ansatz the generalized tortoise coordinate transformation\cite{ZD} 
as follows
\begin{equation}
r_* = r +\frac{1}{2\kappa}\ln[r -r_H(v)] \, , 
~~v_* = v -v_0 \, , \label{trans}
\end{equation}
where $r_H = r_H(v)$ is the location of the event horizon, 
$\kappa$ is an adjustable parameter and is unchanged under 
tortoise transformation. The parameter $v_0$ is an arbitrary 
constant. From formula (\ref{trans}), we can deduce some 
useful relations for the derivatives as follows:
$$ \pr = \left[1 +\frac{1}{\tkr}\right]\spr \, ,
~~\pv = \spv -\frac{r_{H,v}}{\tkr}\spr \, . $$

Now let us consider the asymptotic behavior of $R_1, R_2$ 
near the event horizon. Under the transformation (\ref{trans}), 
Eq. (\ref{sepa}) can be reduced to the following limiting 
form near the event horizon 
\begin{equation}
\spr R_1 = 0 \, , 
~~~~2r_H^2\left[G(r_H) -r_{H,v}\right]\spr R_2 = 0 \, , 
\label{trra} 
\end{equation}
after being taken the  $r \rightarrow r_H(v_0)$ and 
$v \rightarrow v_0$ limits. 

From Eq. (\ref{trra}), we know that $R_1$ is independent 
of $r_*$ and regular on the event horizon. Thus the 
existence condition of a non-trial solution of $R_2$ 
is (as for $r_H \not= 0$)
\begin{equation}
2G(r_H) -2r_{H,v} = 0 \, . \label{loca}
\end{equation}
which  can determine the location of horizon. The event 
horizon equation (\ref{loca}) can be inferred from the 
null hypersurface condition, 
$g^{ij}\partial_i F\partial_j F = 0$, and $F(v,r) = 0$, 
namely $r = r(v)$. A similar procedure applying to the 
null surface equation
\begin{equation}
2\pv F \pr F +2G (\pr F)^2 +\frac{1}{r^2}(\pta F)^2 
+\frac{1}{r^2\sda}(\pvi F)^2 = 0 \, ,
\end{equation}
leads to an equation
\begin{equation}
\Big[2G(r_H) -2r_{H,v}\Big] (\spr F)^2 = 0 \, ,
\end{equation}
resulting in the same event horizon equation due to 
the vanishing of the coefficients in the square bracket. 
As $r_H$ depends on time $v$, the location of the event 
horizon and the shape of the black hole change with time. 

\section{Hawking Temperature}

To investigate the Hawking radiation of spin-$1/2$ particles, 
one need consider the behavior of the second-order form of 
Dirac equation near the event horizon. A straightforward 
calculation gives the second-order radial equation 
\begin{eqnarray} 
&&2r^2\left(G\pdr +\pdvr \right)R_1 +(r^2G)_{,r}\pr R_1 \nonumber\\
&&~~~~~~~~~~~~~ -(\lambda^2 +\mu_0^2r^2)R_1 = -2i\mu_0r^2G R_2 \, , 
\label{sosr+}\\
&&2r^2\left(G\pdr +\pdvr \right)R_2 +3(r^2G)_{,r}\pr R_2 +4r\pv R_2
\nonumber\\ 
&&~~~~~~~~~~~~~ +\left[(r^2G)_{,rr} -(\lambda^2 +\mu_0^2r^2)\right]R_2 
= 2i\mu_0 R_1 \, . \label{sosr-}
\end{eqnarray}

Given the GTCT in Eq. (\ref{trans}) and after some calculations, 
the limiting form of Eqs. (\ref{sosr+},\ref{sosr-}), when $r$ 
approaches $r_H(v_0, \theta_0)$ and $v$ goes to $v_0$, reads
\begin{equation}
\left[\frac{A}{2\kappa} +2G(r_H)\right]\spdr R_1 
+2\spdvr R_1 = 0 \, , \label{ra+}\\
\end{equation}
and
\begin{eqnarray}
&&\left[\frac{A}{2\kappa} +2G(r_H)\right]\spdr R_2 
+2\spdvr R_2 \nonumber\\
&&~~~~+\left[-A +3G_{,r}(r_H) 
+\frac{2G(r_H)}{r_H}\right]\spr R_2 = 0 \, . \label{ra-}
\end{eqnarray}
where we have used relations $2G(r_H) = 2r_{H,v}$ 
and $\spr R_1 = 0$.

With the aid of the event horizon equation (\ref{loca}), 
namely, $2G(r_H) = 2r_{H,v}$, we know that the coefficient 
$A$ is an infinite limit of $0/0$ type. By use of 
the L' H\^{o}spital rule, we get the following result
\begin{equation}
A = \lim_{r \rightarrow r_H(v_0)}\frac{2(G -r_{H,v})}{r -r_H} 
= 2G_{,r}(r_H) \, .
\end{equation}

Now let us select the adjustable parameter $\kappa$ in 
Eqs. (\ref{ra+},\ref{ra-}) such that
\begin{equation}
\frac{A}{2\kappa} +2G(r_H) = \frac{G_{,r}(r_H)}{\kappa} 
+2r_{H,v} \equiv 1 \, ,
\end{equation}
which gives the temperature of the horizon 
\begin{equation}
\kappa =\frac{G_{,r}(r_H)}{1-2G(r_H)}
= \frac{G_{,r}(r_H)}{1-2r_{H,v}} \, . 
\label{temp}
\end{equation}
With such a parameter adjustment, we can reduce 
Eqs. (\ref{ra+},\ref{ra-}) to
\begin{eqnarray}
&&\spdr R_1 +2\spdvr R_1  = 0 \, , 
~~~~\spr R_1 = 0 \, ,\label{wr1} \\
&&\spdr R_2 +2\spdvr R_2 +\left[G_{,r}(r_H) 
+\frac{2G(r_H)}{r_H}\right]\spr R_2  \nonumber\\
&&~~~~ =\spdr R_2 +2\spdvr R_2 +2C\spr R_2 = 0 \, .  
\label{wr2} 
\end{eqnarray}
where $C$ will be regarded as a finite real constant,
$$C =\frac{1}{2} G_{,r}(r_H) +\frac{r_{H,v}}{r_H} \, .$$ 
Eqs. (\ref{wr1},\ref{wr2}) are standard wave equations 
near the horizon, which can be separated by variables
as the following section. 

\section{Thermal Radiation Spectrum}
 
From Eq. (\ref{wr1}), we know that $R_1$ is a constant on 
the event horizon. The solution $R_1 = R_{10}e^{-i\omega v_*}$ 
means that Hawking radiation does not exist for $P_1, Q_2$. 

Now separating variables to Eq. (\ref{wr2}) as  
$R_2 = R_2(r_*)e^{-i\omega v_*}$
and substituting this into equation (\ref{wr2}), one gets
\begin{equation}
 R_2^{\prime\prime} = 2(i\omega -C) R_2^{\prime} \, , 
\end{equation}
The solution is
\begin{equation}
 R_2 = R_{21} e^{2(i\omega -C)r_*} +R_{20} \, . 
\end{equation}

The ingoing wave and the outgoing wave to Eq. (\ref{wr2}) are 
\begin{eqnarray}
&&R_2^{\rm in} = e^{-i\omega v_*} \, , \nonumber\\
&&R_2^{\rm out} = e^{-i\omega v_*} 
e^{2(i\omega -C)r_*} \, ,~~~~~~~ (r > r_H) \, . 
\end{eqnarray}

Near the event horizon, we have 
$$r_* \sim \frac{1}{2\kappa}\ln (r - r_H) \, .$$
Clearly, the outgoing wave $R_2^{\rm out}(r > r_H)$ 
is not analytic at the event horizon $r = r_H$, 
but can be analytically extended from the outside 
of the hole into the inside of the hole through 
the lower complex $r$-plane
$$ (r -r_H) \rightarrow (r_H -r)e^{-i\pi}$$
to
\begin{equation}
\widetilde{R_2^{\rm out}} = e^{-i\omega v_* }
e^{2(i\omega -C)r_*}e^{i\pi C/\kappa}
e^{\pi\omega/\kappa} \, ,~~~~~~(r < r_H) \, . 
\end{equation}

So the relative scattering probability of the 
outgoing wave at the horizon is easily obtained
\begin{equation}
\left|\frac{R_2^{\rm out}}{\widetilde{R_2^{\rm out}}}\right|^2
= e^{-2\pi\omega/\kappa} \, . 
\end{equation}

According to the method of Damour-Ruffini-Sannan's,\cite{DRS} 
the thermal radiation Fermionic spectrum of Dirac particles 
from the event horizon of the hole is given by
\begin{equation} 
\langle {\cal N}_{\omega} \rangle 
= \frac{1}{e^{\omega/T_H} +1} \, , \label{sptr1}
\end{equation} 
with the Hawking temperature $T_H = \frac{\kappa}{2\pi}$ 
\begin{equation}
T_H = \frac{\kappa}{2\pi} 
= \frac{1}{4\pi r_H} \cdot \frac{M r_H   
-\Lambda r_H^4/3 }{M r_H -\Lambda r_H^4/6 } \, .
\end{equation}
It follows that the temperature depends on the time, 
because it is determined by the surface gravity $\kappa$, 
a function of $v$. 

\section{Hawking Radiation
of Klein-Gordon Particles}

To compare the thermal radiation spectrum of electrons
with that of scalar particles, we are now in a position
to investigate the Hawking radiation of Klein-Gordon 
fields. The Klein-Gordon equation $(\Box -\mu^2)\Phi = 0$
for scalar particles with mass $\mu$ in the Vaidya-type 
space-time (1) can be separated by 
$\Phi = R(r)Y_{\ell m}(\theta,\varphi)$
into a radial equation as
\begin{eqnarray} 
&&2r^2\left(G\pdr +\pdvr \right)R +2(r^2G)_{,r}\pr R \nonumber\\ 
&&~~~~ +2r\pv R -[\ell(\ell +1) +\mu^2r^2] R = 0  \, . 
\label{scalar}
\end{eqnarray}
The angular part $Y_{\ell m}(\theta,\varphi)$ is the 
common spherical harmonic function.

Application of a similar prescription as done before 
to Eq. (\ref{scalar}) yields
\begin{eqnarray}
&&\left[\frac{A}{2\kappa} +2G(r_H)\right]\spdr R 
+2\spdvr R \nonumber\\
&&~~~~~~~ +\left[-A +2G_{,r}(r_H) 
+\frac{2G(r_H)}{r_H}\right]\spr R = 0 \, . 
\label{rsra}
\end{eqnarray}
Substituting $A$ into Eq. (\ref{rsra}) and making the
above adjustment of the parameter $\kappa$, we can reduce
the radial part of scalar wave equation to a standard 
form near the horizon
\begin{eqnarray}
&&\spdr R +2\spdvr R +\frac{2G(r_H)}{r_H}\spr R  \nonumber\\
&&~~~~ =\spdr R +2\spdvr R +2C\spr R = 0 \, ,  \label{sw} 
\end{eqnarray}
where $C = \frac{r_{H,v}}{r_H}$. 

Following the measure as done in the preceding section, 
one can easily derive from Eq. (\ref{sw}) the thermal 
radiation Bosonic spectrum of Klein-Gordon particles 
from the event horizon of a Vaidya-type black hole 
\begin{equation} 
\langle {\cal N}_{\omega} \rangle 
= \frac{1}{e^{\omega/T_H} -1} \, , \label{sptr0}
\end{equation} 
The black body radiation spectra (\ref{sptr1}, \ref{sptr0})
demonstrate that no quantum ergosphere effect can appear in
the thermal radiation spectrum of electrons. The difference
between both spectra lies only in the distribution factors 
due to differnt spin statistics.

\section{Conclusions}

Equations (\ref{loca}) and (\ref{temp}) give the location and 
the temperature of event horizon, which depend on the advanced 
time $v$. They are just the same as that obtained in the 
discussion on thermal radiation of Klein-Gordon particles in 
the same space-time. Eqs. (\ref{sptr1},\ref{sptr0}) show, 
respectively, the thermal radiation spectrum of particles with
spin-$1/2, 0$ in a Vaidya-type black hole. These results coincide 
with others. From the thermal spectrum (\ref{sptr1}) of Dirac 
particles, we know that there is not any new interaction 
energy in a Vaidya-type space-time. This manifests that there 
is no new quantum thermal effect called by Li\cite{LZ,LLM} in 
all spherically symmetric black holes. 

In conclusion, we have studied the Hawking radiation of a 
Vaidya-type black hole whose mass changes with time. Our 
results are consistent with others. We have dealt with the 
asymptotic behavior of the separated Dirac equation near the 
event horizon, not only its first-order form but also its 
second-order form. We find that the limiting form of its 
first-order form puts very strong restrict on the Hawking 
radiation, that is, not all components of Dirac spinors but 
$P_2, Q_1$ display the property of thermal radiation. The 
asymmetry of Hawking radiation with respect to the four-component 
Dirac spinors probably originate from the asymmetry of space-times 
in the advanced Eddington-Finkelstein coordinate. This point has 
not been revealed previously. 

In addition, our analysis demonstrates that there was no new 
quantum ergosphere effect in a Vaidya-type space-time as declared 
by Li.\cite{LZ,LLM} This conclusion holds true in any spherically 
symmetric black hole whether it is static or non-static.

\nonumsection{Acknowledgments}
\noindent
S.Q. Wu is indebted to Dr. Jeff Zhao, W. Li for their helps. 
This work is supported in part by the NSFC in China.

\nonumsection{References}

\end{document}